\documentclass[a4paper,11pt]{article}

\usepackage{algorithmic}
\usepackage{algorithm}
\usepackage{graphicx}
\usepackage{citesort}
\usepackage{amssymb,amsmath,amsthm}
\usepackage{pepa}
\usepackage{color}
\usepackage{cite}
\usepackage{fullpage}

\newtheorem{theorem}{Theorem}
\newtheorem{proposition}{Proposition}
\newtheorem{remark}{Remark}
\newtheorem{definition}{Definition}
\newtheorem{model}{Model}

\begin{document}
\thispagestyle{empty}
\title{Numerically  Representing  a Stochastic Process Algebra}
\date{\empty}
\author{Jie Ding\thanks{School of Information Engineering,
Yangzhou University, Yangzhou 225009, P. R. China. Email:
jieding@yzu.edu.cn} and Jane Hillston\thanks{Laboratory for
Foundations of Computer Science, School of Informatics, The
University of Edinburgh, Edinburgh, UK. E-mail:
jane.hillston@ed.ac.uk }}

\maketitle

\begin{quote}
\noindent
{\bf Abstract.} { %\footnotesize
\small The syntactic nature and compositionality characteristic of
stochastic process algebras make models to be easily understood by
human beings, but not convenient for machines as well as people to
directly carry out mathematical analysis and stochastic simulation.
This paper presents a numerical representation schema for the
stochastic process algebra PEPA, which can provide a platform to
directly and conveniently employ a variety of computational
approaches to both qualitatively and quantitatively analyse the
models. Moreover, these approaches developed on the basis of the
schema are demonstrated and discussed. In particular, algorithms for
automatically deriving the schema from a general PEPA model and
simulating the model based on the derived schema to derive
performance measures are presented.}

{\bf Key words: } Numerical Representation;  PEPA; Algorithm

\end{quote}

\newcommand\HI{{\bf I}}

\section{Introduction}

Stochastic process algebras, such as PEPA~\cite{Jane1},
TIPP~\cite{TIPP}, EMPA~\cite{EMPA}, are powerful modelling
formalisms for concurrent systems which have enjoyed considerable
success over the last decade. A stochastic process algebra model is
constructed to approximately and abstractly represent a system
whilst hiding its implementation details. Based on the model,
performance properties of the dynamic behaviour of the system can be
assessed, through some techniques and computational methods. This
process is referred to as the \emph{performance modelling} of the
system, which mainly involves three levels: model construction,
technical computation and performance derivation. In order to derive
performance measures from large scale stochastic process algebra
models, many mathematical tools and approaches have been proposed to
study the models. For instance, a fluid approximation method has
been proposed in~\cite{Jane2} to avoid the state space explosion
problem encountered in the analysis of large scale PEPA models.

\par However, the syntactic nature of stochastic process algebras makes models easily
understood by human beings, but not convenient for
machines/computers (as well as for human beings) to directly employ
these tools and approaches. In addition, the compositionality of the
formalisms allow a model to be locally defined, but the analysis of
the model or the underlying continuous-time Markov chain (CTMC)
carries out in the global manner since it is the whole system rather
than a part of it to be usually interested and considered. The
syntactical and compositional qualities of the stochastic process
algebras, which are advantages in model construction, turn to be
disadvantages in model analysis.

\subsection{Paper contributions}

\par In order to overcome the obstacles in the direct and convenient application
of the mathematical tools, we propose a new numerical representation
schema for the formulism PEPA in this paper. In this schema,
labelled activities are defined to cope with the difference between
actions in PEPA and transitions in the underlying CTMC, so that the
correspondence between them is one-to-one. Activity matrices based
on the labelled activities  are defined to capture structural
information about PEPA models. Moreover, transition rate functions
are proposed to capture the timing information. These concepts
numerically describe and represent a PEPA model, and provide a
platform for conveniently and easily simulating the underlying CTMC,
deriving the fluid approximation, as well as leading to an
underlying Place/Transition (P/T) structure. These definitions are
consistent with the original semantics of PEPA, and a PEPA model can
be recovered from its numerical representation. An algorithm for
automatically deriving the schema from any given PEPA model has been
provided. Some characteristics of this numerical representation are
revealed. For example, using numerical vector forms the exponential
increase of the size of the state space with the number of
components can be reduced to at most a polynomial increase.

\begin{figure}[htbp]
 \begin{center}
 \includegraphics[width=9cm]{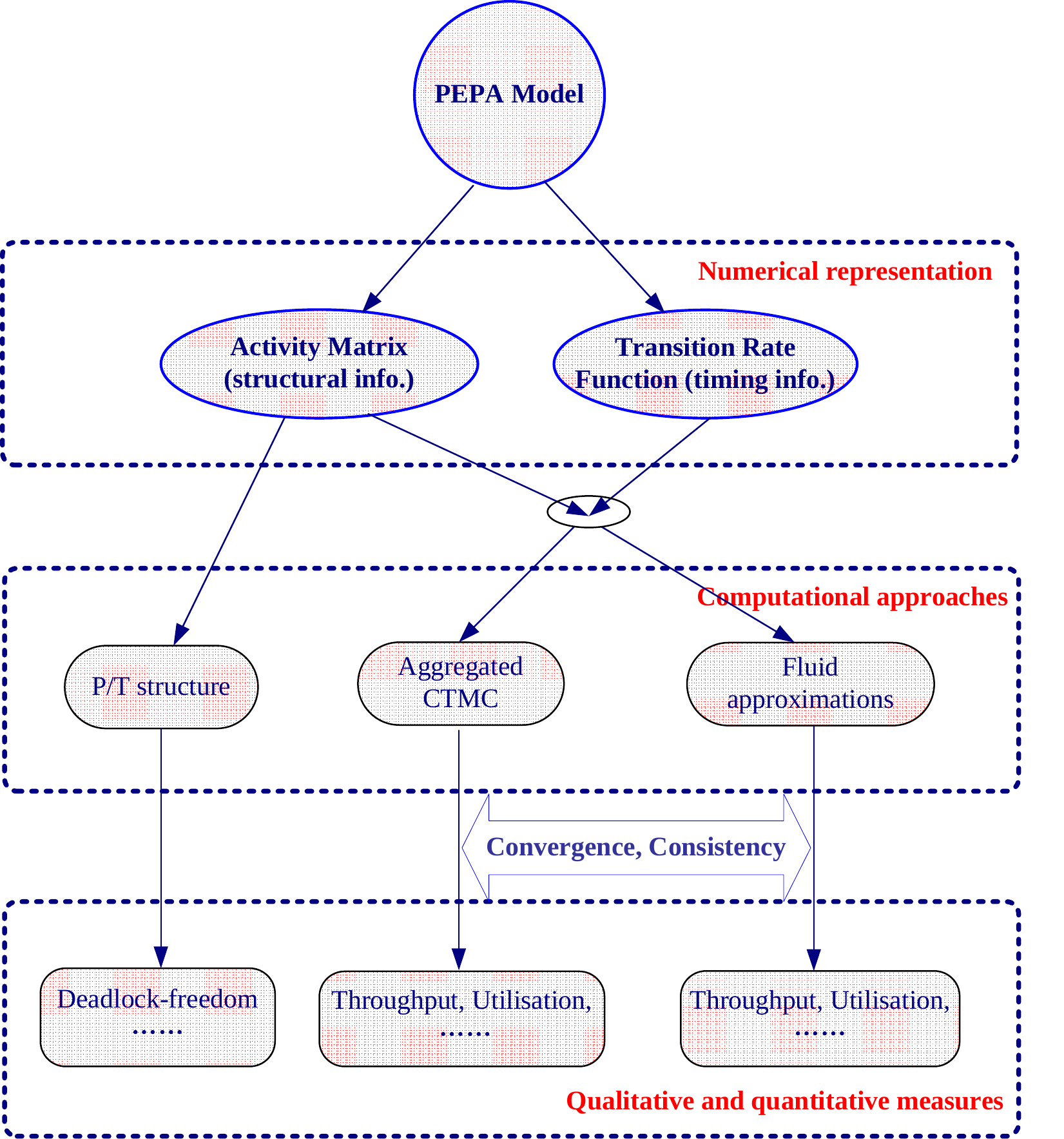}\\
 \end{center} \caption{Three levels of PEPA modelling}\label{fig:Ch1-WorkStructure}
\end{figure}

The benefits of the  schema  embodies the three aspects of
performance modelling, which are illustrated by
Figure~\ref{fig:Ch1-WorkStructure}.  At the first level, the
proposed new schema numerically describes any given PEPA model and
provides a platform  to directly employ a variety of approaches to
analyse the model. These approaches are shown at the second level.
At this level, a fluid approximation method for the quantitative
analysis of PEPA is established, as well as investigated, mainly
with respect to its convergence and the consistency between this
method and the underlying CTMC\@. In addition, a P/T structure-based
approach is revealed, which can be utilised  to qualitatively
analyse the model. At the third level, both qualitative and
quantitative performance measures can be derived from the model
through those approaches.  A stochastic simulation algorithm for the
aggregated CTMC, which is based on the numerical representation
schema, is proposed to obtain general performance metrics in this
paper. As for the other two approaches, related investigation and
analysis were given in~\cite{Structural-Analysis-PEPA}
and~\cite{MircoJie-Fluid-Reward} respectively, which will be briefly
introduced in the next subsection.

\subsection{Related work}

Our work is motivated and stimulated by the pioneering work on the
numerical vector form and activity matrix in~\cite{Jane2}, which was
dedicated to the fluid approximation for PEPA. The P/T structure
underlying each PEPA model, as stated in
Theorem~\ref{theorem:ChStru-P/TinPEPA} in this paper, reveals tight
connections between stochastic process algebras and stochastic Petri
nets. Based on this structure and the theories developed for Petri
nets, several powerful techniques for structural analysis of PEPA
were presented in~\cite{Structural-Analysis-PEPA}, including a
structure-based deadlock-checking method which avoids the state
space explosion problem. In~\cite{Mirco-Fluid-Semantics}, a new
operational semantics was proposed to give a compact symbolic
representation of PEPA models. This semantics extends the
application scope of the fluid approximation of PEPA by
incorporating all the operators of the language and removing earlier
assumptions on the syntactical structure of the models amenable to
this analysis. Moreover, the paper~\cite{MircoJie-Fluid-Reward}
shows how to derive the performance metrics such as action
throughput and capacity utilisaition from the fluid approximation of
a PEPA model.

\subsection{Paper organisation}
The remainder of this paper is structured as follows: Section 2
gives a brief introduction to the PEPA formulism; In Section~3, 4
and 5, we respectively present the three combinators of the
numerical schema, i.e.\ the numerical vector form, labelled activity
and activity matrix, as well as the transition rate function.
Computational approaches for performance derivation that are
developed on the basis of the schema are demonstrated in Section~6.
We finally conclude the paper in Section~7.

\section{Introduction to PEPA}\label{section:PEPA}

PEPA (Performance Evaluation Process Algebra)~\cite{Jane1},
developed by Hillston in the 1990s, is a high-level model
specification language for low-level stochastic models, and
describes a system as an interaction of the components which engage
in activities.  In contrast to classical process algebras,
activities are assumed to have a duration which is a random variable
governed by an exponential distribution. Thus each activity in PEPA
is a pair $(\alpha,r)$ where $\alpha$ is the action type and $r$ is
the activity rate. The language has a small number of combinators,
for which we provide a brief introduction below; the structured
operational semantics can be found in~\cite{Jane1}.  The grammar is
as follows:
\begin{eqnarray*}
  S & ::= & (\alpha, r). S \mid S + S \mid C_S \\
  P & ::= & P \sync{L} P \mid P/L \mid C
\end{eqnarray*}
where $S$ denotes a \emph{sequential component} and $P$ denotes a
\emph{model component} which executes in parallel.  $C$ stands for a
constant which denotes either a sequential component or a model
component as introduced by a definition.  $C_S$ stands for constants
which denote sequential components.  The effect of this syntactic
separation between these types of constants is to constrain legal
PEPA components to be cooperations of sequential processes.

\textbf{Prefix}: The prefix component $(\alpha,r).S$ has a
designated first activity $(\alpha,r)$, which has action type
$\alpha$ and a duration which satisfies exponential distribution
with parameter $r$, and subsequently behaves as $S$.

\textbf{Choice}: The component $S+T$ represents a system which may
behave either as $S$ or as $T$. The activities of both $S$ and $T$
are enabled.  Since each has an associated rate there is a
\emph{race condition} between them and the first to complete is
selected.  This gives rise to an implicit probabilistic choice
between actions dependent of the relative values of their rates.

\textbf{Hiding}: Hiding provides type abstraction, but note that the
duration of the activity is unaffected.  In $P/L$ all activities
whose action types are in $L$ appear as the ``private'' type $\tau$.

\textbf{Cooperation}:$P\sync{L}Q$ denotes cooperation between $P$
and $Q$ over action types in the cooperation set $L$. The cooperands
are forced to synchronise on action types in $L$ while they can
proceed independently and concurrently with other enabled activities
(\emph{individual} activities).  The rate of the synchronised  or
\emph{shared} activity is determined by the slower cooperation (see
\cite{Jane1} for details). We write $P \parallel Q$ as an
abbreviation for $P\sync{L}Q$ when $L = \emptyset$ and $P[N]$ is
used to represent $N$ copies of $P$ in a parallel, i.e.\ $P[3] = P
\parallel P \parallel P$.

\textbf{Constant}: The meaning of a constant is given by a defining
equation such as $A\rmdef P$.  This allows infinite behaviour over
finite states to be defined via mutually recursive definitions.

On the basis of the operational semantic rules (please refer
to~\cite{Jane1} for details),
 a PEPA model may be regarded as a labelled multi-transition system
$$\left(\mathcal{C},\mathcal{A}ct,
\left\{\mathop{\longrightarrow}\limits^{(\alpha,r)}|(\alpha,r)\in
\mathcal{A}ct\right\}\right)$$ where $\mathcal{C}$ is the set of
components, $\mathcal{A}ct$ is the set of activities and the
multi-relation $\mathop{\longrightarrow}\limits^{(\alpha,r)}$ is
given by the rules. If a component $P$ behaves as $Q$ after it
completes activity $(\alpha, r)$, then denote the transition as
$P\mathop{\longrightarrow}\limits^{(\alpha,r)}Q$.

The memoryless property of the exponential distribution, which is
satisfied by the durations of all activities, means that the
stochastic process underlying the labelled transition system has the
Markov property. Hence the underlying stochastic process is a
CTMC\@. Note that in this representation the states of the system
are the syntactic terms derived by the operational semantics.  Once
constructed the CTMC can be used to find steady-state or transient
probability distributions from which quantitative performance  can
be derived.

\section{Numerical Vector Form}\label{section:Ch3-NumVectorForm}

The usual state representation in PEPA models is in terms of the
syntactic forms of the model expression.  When a large number of
repeated components are involved in a system, the state space of the
CTMC underling the model can be large. This is mainly because each
copy of the same type of component is considered to be distinct,
resulting in distinct Markovian states. The multiple states within
the model that exhibit the same behaviour can be aggregated to
reduce the size of the state space as shown by Gilmore \emph{et
al.}~\cite{Ribaudo-AggregatingAlgorithm} using the technique based
on a vector form. The CTMC is therefore constructed in terms of
equivalence classes of syntactic terms. ``At the heart of this
technique is the use of a canonical state vector to capture the
syntactic form of a model expression", as indicated in~\cite{Jane2}.
% ``if two states have the same canonical
%state vector they are equivalent and need not be distinguished in
%the aggregated derivation graph''.
Rather than the canonical representation style, an alternative
numerical vector form was proposed by Hillston in~\cite{Jane2} for
capturing the state information of models with repeated components.
In the numerical vector form, there is one entry for each local
derivative of each type of component in the model. The entries in
the vector are the number of components currently exhibiting this
local derivative,  no longer syntactic terms representing the local
derivative of the sequential component. Following~\cite{Jane2},
hereafter the term \emph{local derivative} refers to the local state
of a single sequential component.

\begin{definition} (\textbf{{Numerical Vector Form}\cite{Jane2}})\label{def:Ch3-NumVectorForm}.
For an arbitrary PEPA model $\mathcal{M}$ with $n$ component types
${C}_i, i=1,2,\cdots,n$, each with $d_i$ distinct local derivatives,
the numerical vector form of $\mathcal{M}$,
$\mathbf{m}(\mathcal{M})$, is a vector with $d=\sum_{i=1}^nd_i$
entries. The entry $\mathbf{m}[{C}_{i_j}]$ records how many
instances of the $j$th local derivative of component type ${C}_i$
are exhibited in the current state.
\end{definition}

By adopting this model-aggregation technique, the number of the
states of the system can be reduced to only increase (at most)
polynomially with the number of instances of the components.
According to Definition~\ref{def:Ch3-NumVectorForm},
$\mathbf{m}({C}_{i_j})\geq 0$ for each ${C}_{i_j}$. At any time,
each sequential component stays in one and only one local
derivative. So the sum of $\mathbf{m}({C}_i)$, i.e.
$\sum_{j=1}^{d_i}\mathbf{m}[{C}_{i_{j}}]$, specifies the population
of ${C}_{i}$ in the system. Notice that $\mathbf{m}({C}_i)$
satisfies
\begin{equation}\label{eq:Ch3-NumVectCondition1}
  \left\{\begin{array}{c}
           \mathbf{m}[{C}_{i_1}],\mathbf{m}[{C}_{i_2}],
\cdots,\mathbf{m}[{C}_{i_{d_i}}]\in \mathbb{Z}^{+},\\
           \sum_{j=1}^d\mathbf{m}[{C}_{i_j}]=M_i.
         \end{array}\right.
\end{equation}
 Then according to the well-known combinatorial formula
(Theorem~3.5.1 in~\cite{Brualdi-IntroCombinatorics}), there are
$\left(
                     \begin{array}{c}
                       M_i+d_i-1 \\
                       d_i-1 \\
                     \end{array}
                   \right)$
 solutions, i.e. $\left(
                     \begin{array}{c}
                       M_i+d_i-1 \\
                       d_i-1 \\
                     \end{array}
                   \right)$ states in terms of ${C}_i$
in the system. But the possible synchronisations in the PEPA model
have not been taken into account in the restrictions
(\ref{eq:Ch3-NumVectCondition1}) and thus the current restrictions
may allow extra freedom for the solutions, so the given number
$\left(
                     \begin{array}{c}
                       M_i+d_i-1 \\
                       d_i-1 \\
                     \end{array}
                   \right)$
is an upper bound of the exact number of the states in terms of
${C}_i$. Notice that
\begin{equation*}
\begin{split}
\left( \begin{array}{c}
         M_i+d_i-1 \\
              d_i-1\\
         \end{array}\right)
         %=&\frac{(M_i+d_i-1)(M_i+d_i-2)\cdots (M_i+1)}{(d_i-1)!}\\
         %=&\frac{(M_i+d_i-1)!}{(d_i-1)!M_i!} \leq \left(M_i+d_i-1\right)^{d_i-1}.\\
         =&\frac{(M_i+d_i-1)!}{(d_i-1)!M_i!} \\
         \leq&  \left(M_i+d_i-1\right)^{d_i-1}.
\end{split}
\end{equation*}
Therefore, it is easy to verify the following
\begin{proposition}\label{proposition:Ch3-NumericalVector}
Consider a system comprising $n$ types of component, namely ${C}_1,
\mathcal{C}_2, \cdots, {C}_n$, with $M_i$ copies of the component of
type ${C}_i$ in the system, where ${C}_i$ has $d_i$ local
derivatives, for $i=1,2,\cdots,n$. Then the size of the state space
of the system is at most $$
   \prod_{i=1}^n     \left(
                             \begin{array}{c}
                               M_i+d_i-1 \\
                                d_i-1\\
                             \end{array}
                           \right)\leq \prod_{i=1}^n
                           \left(M_i+d_i-1\right)^{d_i-1}.
$$

\end{proposition}
The upper bound given in
Proposition~\ref{proposition:Ch3-NumericalVector} guarantees that
the size of the state space increases at most \emph{polynomially}
with the number of instances of the components. Consider the
following PEPA model.
\begin{model}\label{model:Uer-Provider}%\textbf{PEPA Model of User-Provider System}\\
\begin{equation*}
\begin{split}
User_1 \rmdef &(task_1, a).User_2\\
User_2 \rmdef &(task_2, b).User_1\\
Sever_1 \rmdef &(task_1, a).Sever_2\\
Sever_2 \rmdef &(reset, d).Sever_1\\
{User_1[M]}& \sync{\{task_1\}}
  {Sever_1[N]}.
\end{split}
\end{equation*}
\end{model}
\begin{figure}[htbp]
 \begin{center} \includegraphics[width=9cm]{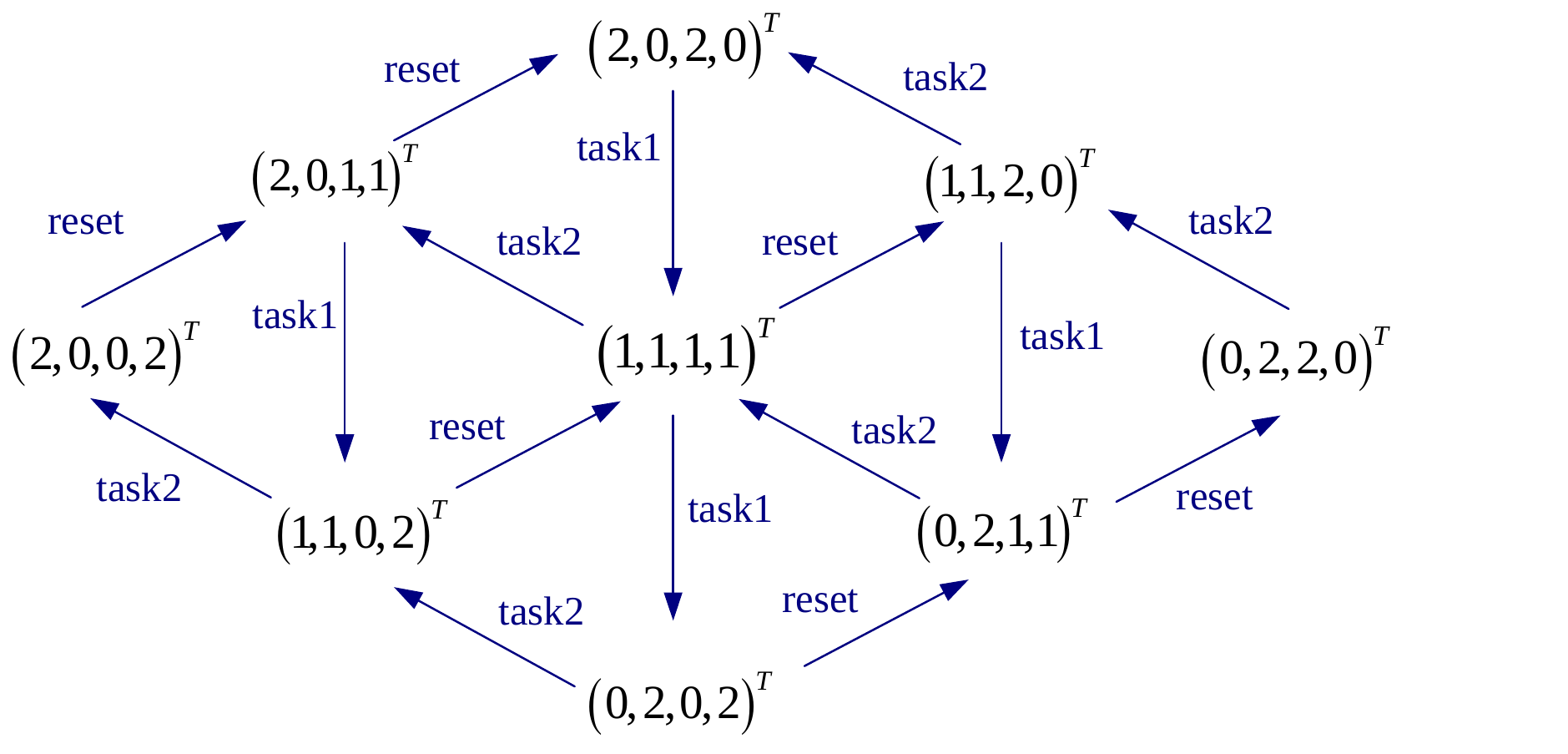}\\
 \end{center} \caption[Transition between states (Model~\ref{model:Uer-Provider})]
 {Transition between states (a revised version of the one in~\cite{Jane2})}
 \label{fig:Ch3-UserProviderTransition}
\end{figure}
According to the semantics of PEPA originally defined
in~\cite{Jane1}, the size of the state space of the CTMC underlying
Model~\ref{model:Uer-Provider} is $2^{M+N}$, which increases
exponentially with the numbers of the users and severs in the
system. According to Definition~\ref{def:Ch3-NumVectorForm}, the
system vector $\mathbf{m}$ has four entries representing the
instances of components in the total four local derivatives, that is
\begin{equation*}
\begin{split}
     \mathbf{m}=\left(\mathbf{m}[User_1],\mathbf{m}[User_2],\mathbf{m}[Sever_1],\mathbf{m}[Sever_2]\right)^T.
\end{split}
\end{equation*}
Let $M=N=2$, then the system equation of
Model~\ref{model:Uer-Provider} determines the starting state
$(2,0,2,0)$. By enabling activities or transitions, all reachable
system states can be manifested, see
Figure~\ref{fig:Ch3-UserProviderTransition}. The size of the state
space is nine. The upper bound of the size given by
Proposition~\ref{proposition:Ch3-NumericalVector}, $\left(
\begin{array}{c}
                               M+2-1 \\
                                2-1\\
                             \end{array}
                           \right)\left(
                             \begin{array}{c}
                               N+2-1 \\
                                2-1\\
                             \end{array}
                           \right)$ or $(M+1)\times
(N+1)$, is nine, coinciding with the size of the state space.
% It is
%a significant improvement to reduce the size of the state space from
%$2^{M}\times 2^N$ to  $(M+1)\times (N+1)$, without relevant
%information and accuracy loss.
The bound given in Proposition~\ref{proposition:Ch3-NumericalVector}
is sharp and can be hit in some situations.

\section{Labelled Activity and Activity
Matrix}\label{section:Ch3-ActivityMatrix}

In the PEPA language, the transition is embodied in the syntactical
definition of activities, in the context of sequential components.
Since the consideration is in terms of the whole system rather than
sequential components, the transition between these system states
should be defined and represented. This section presents a numerical
representation for the transitions between  system states and
demonstrates how to derive this representation from a general PEPA
model.

If a system vector changes into another vector after firing an
activity, then the difference between these two vectors manifests
the transition corresponding to this activity. Obviously, the
difference is in numerical forms since all states are numerical
vectors. Consider Model~\ref{model:Uer-Provider} and its transition
diagram in Figure~\ref{fig:Ch3-UserProviderTransition}. Each
activity in the model corresponds to a  vector, called the
\emph{transition vector}. For example, $task_1$ corresponds to the
transition vector $l^{task_1}=(-1,1,-1,1)^T$. That is, the derived
state vector by firing $task_1$ from a state, can be represented by
the sum of $l^{task_1}$ and the state enabling $task_1$. For
instance, $(2,0,2,0)^T+l^{task_1}=(1,1,1,1)^T$ illustrates that
$\mathbf{s}_1=(2,0,2,0)^T$ transitions into
$\mathbf{s}_2=(1,1,1,1)^T$ after enabling $task_1$.

\begin{table}[htbp]
\begin{center}\caption{Transition vectors form an activity
   matrix}\label{fig:Ch3-TVectorFormActivityMatrix}
\begin{tabular}{c| c |c| c |c}
  %\hline\hline
             &   $l^{task_1}$ &   $l^{task_2}$ &   $l^{reset}$ & \\\hline
  $User_1$& $-1$ & 1  & 0   & \\%$x_1$             \\
  $User_2$& 1  & $-1$  & 0   & \\%$x_2$             \\
  $Sever_1$ & $-1$  & 0  & 1   & \\%$x_3$             \\
  $Sever_2$ & 1  & 0  & $-1$   & \\%$x_4$             \\
  \end{tabular}
   \end{center}
\end{table}

\par  Similarly, $task_2$ corresponds to $l^{task_2}=(1,-1,0,0)^T$  while $reset$ corresponds to
$l^{reset}=(0,0,-1,1)$. The three transition vectors form a matrix,
called the \emph{activity matrix}, see
Table~\ref{fig:Ch3-TVectorFormActivityMatrix}. Each activity in the
model is represented by a transition vector --- a column of the
activity matrix, and each column expresses an activity. So the
activity matrix is essentially indicating both an injection and a
surjection from syntactic to numerical representation of the
transition between system states. The  concept of the activity
matrix for PEPA was first proposed by Hillston
in~\cite{Hillston-ODE-first,Jane2}. However, the original definition
cannot fully reflect the representation mapping considered here.
This is due to the fact of that the original definition is
local-derivative-centric rather than transition centric. This
results in some limitations for more general applications. For
example, for some PEPA models (e.g.\ Model~\ref{model:Ch3-P3Q2} in
the following context), some columns of the originally defined
matrix cannot be taken as transition vectors so that this definition
cannot fully reflect the PEPA semantics in some circumstances. In
the following, a modified definition of the activity matrix is
given. The new definition is activity- or transition-centric, which
brings the benefit that each transition is represented by a column
of the matrix and vice versa.

\begin{model}\label{model:Ch3-P3Q2}
\begin{equation*}
\begin{split}
P_1 \rmdef &(\alpha, r_{\alpha}').P_2+(\alpha, r_{\alpha}'').P_3\\
P_2 \rmdef &(\beta, r_{\beta}).P_1+(\beta, r_{\beta}').P_3\\
P_3 \rmdef &(\gamma, r_{\gamma}).P_1\\
Q_1 \rmdef &(\alpha, r_{\alpha}).Q_2\\
Q_2 \rmdef &(\gamma, r_{\gamma}').Q_1\\
{P_1[A]}& \sync{\{\alpha\}}
  {Q_1[B]}.
\end{split}
\end{equation*}
\end{model}

\begin{figure}[htbp]
 \begin{center} \includegraphics[width=8.5cm]{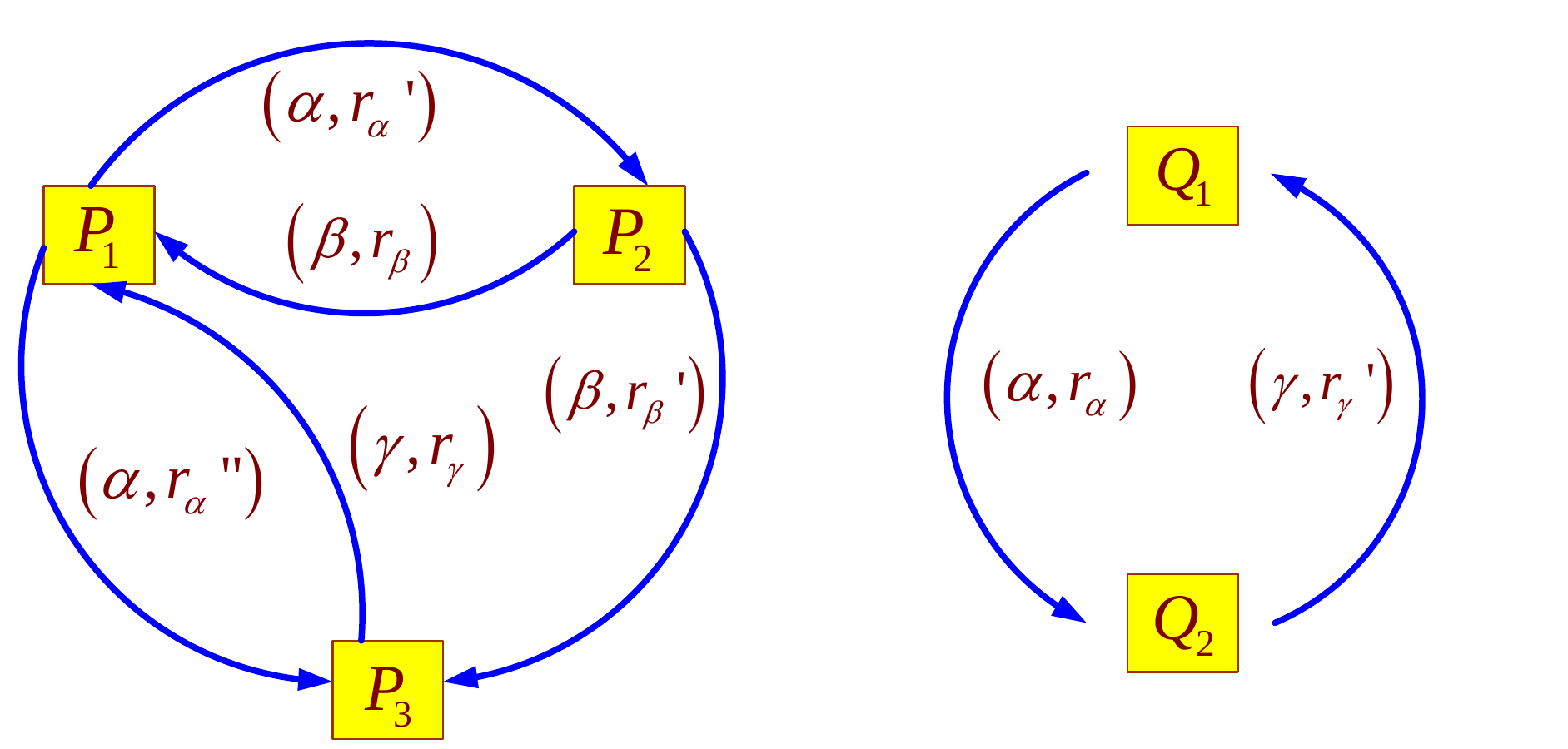}\\
 \end{center} \caption[Transition diagram of Model~\ref{model:Ch3-P3Q2}]
 {Transition diagram of Model~\ref{model:Ch3-P3Q2}}
 \label{fig:Ch3-Diagram-Model2}
\end{figure}

\begin{table}[htbp]
\begin{center}
\caption{Originally defined activity matrix of
Model~\ref{model:Ch3-P3Q2}}\label{table:Matrix-Ch3-P3Q2}
\begin{tabular}{c| c|c |c}
  %\hline\hline
           &   $\alpha$ &   $\beta$ & $\gamma$
             \\\hline
  $P_1$& $-1$  & $1$  &  1   \\
  $P_2$& $1$   & $-1$ &  0   \\
  $P_3$& $1$   & $1$  &  $-1$   \\
  $Q_1$& $-1$  & 0    & $1$  \\
  $Q_2$& $1$   & $0$  & $-1$   \\
  \end{tabular}
\end{center}
\end{table}

\par Let us consider a PEPA model, i.e.\ the following Model~\ref{model:Ch3-P3Q2},
in which there are multiple choices after firing some activities. In
this model, firing $\alpha$ in the component $P$ may lead to two
possible local derivatives: $P_2$ and $P_3$, while firing $\beta$
may lead to $P_1$ and $P_3$. In addition, firing $\gamma$ may lead
to $P_1,Q_1$. However, only one derivative can be chosen after each
firing of an activity, according to the semantics of PEPA\@. But the
original definition of activity matrix cannot clearly reflect this
point. See the activity matrix of Model~\ref{model:Ch3-P3Q2} given
in Table~\ref{table:Matrix-Ch3-P3Q2}. Moreover, the individual
activity $\gamma$ in this table, which can be enabled by both $P_3$
and $Q_2$, may be confused as a shared activity.

\par In order to better reflect the semantics of PEPA, we modify the
definition of the activity matrix in this way: if there are $m$
possible outputs, namely $\{R_1,R_2,\cdots,R_m\}$, after firing
either an individual or a shared activity $l$, then $l$  is
``split'' into $m$ labelled $l$s: $l^{w_1}, l^{w_2}, \cdots,
l^{w_m}$. Here  $\{w_i\}_{i=1}^m$ are $m$ distinct labels,
corresponding to $\{R_i\}_{i=1}^m$ respectively. Each $l^{w_i}$ can
only lead to a unique output $R_i$. Here there are no new activities
created, since we just attach labels to the activity to distinguish
the outputs of firing this activity. The modified activity matrix
clearly reflects that only one, not two or more, result can be
obtained from firing $l$. And thus, each $l^{w_i}$ can represent a
transition vector.

\par For example, see the modified activity matrix of Model~\ref{model:Ch3-P3Q2} in
Table~\ref{table:ModifiedMatrix-Ch3-P3Q2}. In this activity matrix,
the individual activity $\gamma$ has different ``names'' for
different component types, so that it is not confused with a shared
activity. Another activity $\beta$, is labelled as
$\beta^{P_2\rightarrow P_1}$ and $\beta^{P_2\rightarrow P_3}$, to
respectively reflect the corresponding two choices. In this table,
the activity $\alpha$ is also split and attached with labels.

\begin{table}[htbp]
\begin{center}
\caption{Modified activity matrix of
Model~\ref{model:Ch3-P3Q2}}\label{table:ModifiedMatrix-Ch3-P3Q2}
\begin{tabular}[t]{c|c |c|c|c|c|c}
  %\hline\hline
             &
              \rotatebox[origin=c]{90}{$\alpha^{(P_1\rightarrow P_2,Q_1\rightarrow
             Q_2)}$}
             & \rotatebox[origin=c]{90}{$\alpha^{(P_1\rightarrow P_3,Q_1\rightarrow
             Q_2)}$}
             &  %\rotatebox[origin=c]{90}
             {$\beta^{P_2\rightarrow P_1}$}
             & %\rotatebox[origin=c]{90}
               {$\beta^{P_2\rightarrow P_3}$}
             &  %\rotatebox[origin=c]{90}
                 {$\gamma^{P_3\rightarrow P_1}$}
             & %\rotatebox[origin=c]{90}
              {$\gamma^{Q_2\rightarrow Q_1}$}
             \\\hline
  $P_1$& $-1$ & $-1$&$1$   &$0$  & 1    & 0   \\
  $P_2$& $1$  & 0   &$-1$  &$-1$ & $0$  & 0   \\
  $P_3$& $0$  & 1   &  0   & 1   & $-1$ & 0    \\
 $Q_1$& $-1$  & $-1$&  0   & 0   & 0    & 1    \\
 $Q_2$& $1$   & 1   &  0   & 0   & $0$  & $-1$  \\
 \end{tabular}
\end{center}
\end{table}

\par Before giving the modified definition of activity matrix for any
general PEPA model, the pre and post sets for an activity are first
defined. For convenience, throughout this paper any  transition
$U\stackrel{(l,r)}{\longrightarrow}V$ defined in the PEPA models may
be rewritten as $U\stackrel{(l,r_l^{U\rightarrow
V})}{\longrightarrow}V$, or just $U\stackrel{l}{\longrightarrow}V$
if the rate is not being considered, where $U$ and $V$ are two local
derivatives.

\begin{definition}(\textbf{Pre and post local
derivative})\label{def:Ch3-pre-post-derivative}
\begin{enumerate}
  \item If a local derivative $U$ can enable an activity $l$, that
is $U\mathop{\longrightarrow}\limits^{l}\cdot$, then $U$ is called a
\emph{pre local derivative} of $l$. The set of all pre local
derivatives of $l$ is denoted by $\mathrm{pre}(l)$, called the
\emph{pre set} of $l$.

  \item  If $V$ is a
local derivative obtained by firing an activity $l$, i.e.
$\cdot\mathop{\longrightarrow}\limits^{l}V$, then $V$ is called a
\emph{post local derivative} of $l$. The set of all post local
derivatives is denoted by $\mathrm{post}(l)$, called the \emph{post
set} of $l$.

  \item  The set of all the local derivatives derived from
$U$ by firing $l$, i.e.
$$
      \mathrm{post}(U,l)=\{V\mid U\stackrel{l}{\longrightarrow}V\},
$$
is called the \emph{post set of $l$ from
  $U$}.
\end{enumerate}

\end{definition}

Obviously, if $l$ has only one pre local derivative, i.e.
$\#\mathrm{pre}(l)=1$, then $l$ is an individual activity, like
$\beta$ in Model~\ref{model:Ch3-P3Q2}, whereafter the notation $\#
A$ is defined as the cardinality of the set $A$, i.e.\ the number of
elements of $A$. But $l$ being individual does not imply
$\#\mathrm{pre}(l)=1$, see $\gamma$ for instance. If $l$ is shared,
then $\#\mathrm{pre}(l)>1$, for example, see
$\#\mathrm{pre}(\alpha)=\#\{P_1, Q_1\}=2$. For a shared activity $l$
with $\mathrm{pre}(l)=k$, there are $k$ local derivatives that can
enable this activity, each of them belonging to a distinct component
type.  The obtained local derivatives are in the set
$\mathrm{post}(\mathrm{pre}(l)[i],l)$, where $\mathrm{pre}(l)[i]$ is
the $i$-th pre local derivative of $l$. But only one of them can be
chosen after $l$ is fired from $\mathrm{pre}(l)[i]$. Since for the
component type, namely $i$ or $\mathcal{C}_i$, there are
$\#\mathrm{post}(\mathrm{pre}(l)[i],l)$ outputs, so the total number
of the distinct transitions for the whole system is $
\prod_{i=1}^k\#\mathrm{post}(\mathrm{pre}(l)[i],l). $ That is, there
are $\prod_{i=1}^k\#\mathrm{post}(\mathrm{pre}(l)[i],l)$ possible
results but only one of them can be chosen by the system after the
shared activity $l$ is fired. In other words, to distinguish these
possible transitions, we need
$\prod_{i=1}^k\#\mathrm{post}(\mathrm{pre}(l)[i],l)$ different
labels. Here are the readily accessible labels:
$$
(\mathrm{pre}(l)[1]\rightarrow V_1,\mathrm{pre}(l)[2]\rightarrow
V_2,\cdots,\mathrm{pre}(l)[k]\rightarrow V_k),
$$
where $V_i\in\mathrm{post}(\mathrm{pre}(l)[i],l)$. Obviously, for
each vector $(V_1,V_2,\cdots, V_k)$ in $
\mathrm{post}(\mathrm{pre}(l)[1],l)\times
\mathrm{post}(\mathrm{pre}(l)[2],l)\times\cdots\times
\mathrm{post}(\mathrm{pre}(l)[k],l),$
 the labelled activity
$l^{(\mathrm{pre}(l)[1]\rightarrow V_1,\mathrm{pre}(l)[2]\rightarrow
V_2,\cdots,\mathrm{pre}(l)[k]\rightarrow V_k)}$ represents  a
distinct transition. For example, $\alpha$ in
Model~\ref{model:Ch3-P3Q2} can be labelled as
$\alpha^{(P_1\rightarrow P_2,Q_1\rightarrow Q_2)}$ and
$\alpha^{(P_1\rightarrow P_3,Q_1\rightarrow Q_2)}$.

\par For an individual activity $l$,  things are rather
 simple and easy: for  $U\in \mathrm{pre}(l)$, $l$ can be labelled
 as $l^{U\rightarrow \mathrm{post}(U,l)[1]}$, $l^{U\rightarrow
\mathrm{post}(U,l)[2]}$, $l^{U\rightarrow \mathrm{post}(U,l)[k_U]},$
where $k_U=\#\mathrm{post}(U,l)$. Varying $U\in \mathrm{pre}(l)$,
there are $ \sum_{U\in \mathrm{pre}(l)}\#\mathrm{post}(U,l) $ labels
needed to distinguish the possible transitions. See $
\beta^{P_2\rightarrow P_1},\beta^{P_2\rightarrow P_3},
\gamma^{P_3\rightarrow P_1},\gamma^{Q_2\rightarrow Q_1} $ in
Model~\ref{model:Ch3-P3Q2} for instance. Now we give the formal
definition.

\begin{definition}\label{definition:LabelledAcitvity}(\textbf{Labelled
Activity}).

\begin{enumerate}
  \item For any individual activity $l$, for
each  $U\in \mathrm{pre}(l), V\in \mathrm{post}(U,l)$, label $l$ as
$l^{U\rightarrow V}$.

 \item For a shared activity $l$, for each
\begin{equation*}
\begin{split}
 (V_1,V_2,\cdots,V_k)\in
 \mathrm{post}(\mathrm{pre}(l)[1],l)\times
\mathrm{post}(\mathrm{pre}(l)[2],l)\times\cdots \times
\mathrm{post}(\mathrm{pre}(l)[k],l),
\end{split}
\end{equation*}
label $l$ as $l^{w}$, where
\begin{equation*}
\begin{split}
 w=(\mathrm{pre}(l)[1]\rightarrow V_1,\mathrm{pre}(l)[2]\rightarrow
V_2,\cdots, \mathrm{pre}(l)[k]\rightarrow V_k).
\end{split}
\end{equation*}

\end{enumerate}
Each $l^{U\rightarrow V}$ or $l^{w}$ is called a \emph{labelled
activity}. The set of all labelled activities  is denoted by
$\mathcal{A}_{\mathrm{label}}$. For the above labelled activities
$l^{U\rightarrow V}$ and $l^{w}$, their respective pre and post sets
are defined as
$$\mathrm{pre}(l^{U\rightarrow V})=\{U\},\;\mathrm{post}(l^{U\rightarrow
V})=\{V\},$$
$$\mathrm{pre}(l^{w})=\mathrm{pre}(l),\;\mathrm{post}(l^{w})=\{V_1,V_2,\cdots,V_k\}.$$
\end{definition}

\par According to Definition~\ref{definition:LabelledAcitvity}, each
$l^{U\rightarrow V}$ or $l^{w}$ can only lead to a unique output. No
new activities are created, since labels are only attached to the
activity to distinguish the results after this activity is fired.

\par The impact of labelled activities on local derivatives can be recorded in
 a matrix form, as defined below.

\begin{definition}\label{definition: JieActivityMatrix}(\textbf{{Activity
Matrix}}). For a model with $N_{\mathcal{A}_{\mathrm{label }}}$
labelled activities and $N_\mathcal{D}$ distinct local derivatives,
the activity matrix $\mathbf{C}$ is an $N_\mathcal{D}\times
N_{\mathcal{A}_{\mathrm{label}}}$ matrix, and the entries are
defined as follows
\begin{equation*}
     \mathbf{C}(U_i,l_j)=\left\{
             \begin{array}{cl}
               +1 & \mbox{if $U_i\in \mathrm{post}(l_j)$}\\
               -1 & \mbox{if $U_i\in \mathrm{pre}(l_j)$}\\
               0 & \mathrm{otherwise}
             \end{array}
     \right.
\end{equation*}
where $l_j$ is a labelled activity. The pre activity matrix
$\mathbf{C^{pre}}$ and post activity matrix $\mathbf{C^{post}}$  are
defined as
\begin{equation*}
     \mathbf{C^{Pre}}(P_i,\alpha_j)=\left\{
             \begin{array}{ll}
               +1 \hspace*{3mm}  & \mbox{if } \;\; P_i\in \mathrm{pre}(\alpha_j) \\
                0 & \mbox{otherwise}.
             \end{array}
     \right.,
\end{equation*}
\begin{equation*}
     \mathbf{C^{Post}}(P_i,\alpha_j)=\left\{
             \begin{array}{ll}
               +1 \hspace*{3mm} & \mbox{if } \;\;P_i\in \mathrm{post}(\alpha_j)\\
                0 & \mbox{otherwise}.
             \end{array}
     \right.
\end{equation*}
\end{definition}

The modified activity matrix captures all the structural
information, including the information about choices and
synchronisations, of a given PEPA model. From each row of the
matrix, which corresponds to each local derivative, we can know
which activities this local derivative can enable and after which
activities are fired this local derivative can be derived. From the
perspective of the columns, the number of ``$-1$''s in a column
tells whether the corresponding activity is synchronised or not.
Only one ``$-1$'' means that this transition corresponds to an
individual activity. The locations of ``$-1$'' and ``$1$'' indicate
which local derivatives can enable the activity and what the derived
local derivatives are, i.e.\ the pre and post local derivatives. In
addition, the numbers of ``$-1$''s and ``$1$''s in each column are
the same, because any transition in any component type corresponds
to a unique pair of pre and post local derivatives. In fact, all
this information is also stored in the labels of the activities.
Therefore, with the transition rate functions defined in the next
section to capture the timing information, a given PEPA model can be
recovered from its activity matrix.

Moreover, the pre and post activity matrix indicate the local
derivatives which can fire a labelled activity and the derived local
derivative after firing a labelled activity respectively.
%Clearly, the pre activity matrix indicates the pre local derivatives
%for each labelled activity, i.e. the local derivatives which can
%fire this activity. The post activity matrix indicates the post
%local derivatives, i.e. the derived local derivatives after firing
%an activity.
 The modified activity matrix equals the difference
between the pre and post activity matrices, i.e.\
$\mathbf{C=C^{Post}-C^{Pre}}$. Hereafter the terminology of
\emph{activity matrix} refers to the one in
Definition~\ref{definition: JieActivityMatrix}. This definition
embodies the transition or operation rule of a given PEPA model,
with the exception of timing information. For a given PEPA model,
each transition of the system results from the firing of an
activity. Each optional result after enabling this activity
corresponds to a relevant labelled activity, that is, corresponds to
a column of the activity matrix. Conversely, each column of the
activity matrix corresponding to a labelled activity, represents an
activity and the chosen derived result after this activity is fired.
So each column corresponds to a system transition. Therefore, we
have the following proposition, which specifies the correspondence
between system transitions and the columns of the activity matrix.
\begin{proposition}\label{proposition:Ch3-TransVSActivityMatrix}
Each column of the activity matrix corresponds to a system
transition and each transition can be represented by a column of the
activity matrix.
\end{proposition}

%
%
%
%{\color{red}: when does a transition happen or occur?}

\section{Transition Rate
Function}\label{section:Ch3-TransitionRateFunction}

The structural information of any general PEPA model is captured in
the activity matrix, which is constituted by all transition vectors.
However, the duration of each transition has not yet been specified.
This section defines transition rate functions for transition
vectors or labelled activities to capture the timing information of
PEPA models.

\subsection{Model~\ref{model:Ch3-P3Q2}
continued}\label{section:Ch3-FrontDiscussionRateFunct}

\par   Let us start from
Model~\ref{model:Ch3-P3Q2} again. As
Table~\ref{table:ModifiedMatrix-Ch3-P3Q2} shows, activity $\gamma$
in Model~\ref{model:Ch3-P3Q2} is labelled as $\gamma^{P_3\rightarrow
P_1}$ and $\gamma^{Q_2\rightarrow Q_1}$. For $\gamma^{P_3\rightarrow
P_1}$, there are $\mathbf{x}[P_3]$ instances of the component type
$P$ in the local derivative $P_3$ in state $\mathbf{x}$, each
enabling the individual activity concurrently with the rate
$r_{\gamma}$. So the rate of $\gamma^{P_3\rightarrow P_1}$ in state
$\mathbf{x}$ is $f(\mathbf{x},\gamma^{P_3\rightarrow
P_1})=r_{\gamma}\mathbf{x}[P_3]$. Similarly, the rate for
$\gamma^{Q_2\rightarrow Q_1}$ in state $\mathbf{x}$ is
$r_{\gamma'}\mathbf{x}[Q_2]$. This is consistent with the definition
of apparent rate in PEPA, which states that if there are $N$
replicated instances of a component enabling a transition $(l,r)$,
the apparent rate of the activity will be $r\times N$.

\par In Model~\ref{model:Ch3-P3Q2} activity $\beta$ is labelled as
$\beta^{P_2\rightarrow P_1}$ and $\beta^{P_2\rightarrow P_3}$, to
respectively reflect the corresponding two choices. According to the
model definition, there is a flux of $r_{\beta}\mathbf{x}(P_2)$ into
$P_1$ from $P_2$ after firing $\beta$ in state $\mathbf{x}$. So the
transition rate function is defined as
$f(\mathbf{x},\beta^{P_2\rightarrow P_1})=r_{\beta}\mathbf{x}[P_2]$.
Similarly, we can define $f(\mathbf{x},\beta^{P_2\rightarrow
P_3})=r_\beta' \mathbf{x}[P_2]$. These rate functions can be defined
or interpreted in an alternative way. In state $\mathbf{x}$, there
are $\mathbf{x}[P_2]$ instances that can fire $\beta$. So the
apparent rate of $\beta$ is $(r_\beta+r_\beta')\mathbf{x}[P_2]$. By
the semantics of PEPA, the probabilities of choosing the outputs are
$\frac{r_\beta}{r_\beta+r_\beta'}$ and
$\frac{r_\beta'}{r_\beta+r_\beta'}$ respectively. So the rate of the
transition $\beta^{P_2\rightarrow P_1}$ is
\begin{equation}\label{eq:Ch3-f(x,beta)-1}
f(\mathbf{x},\beta^{P_2\rightarrow
P_1})=\frac{r_\beta}{r_\beta+r_\beta'}
(r_\beta+r_\beta')\mathbf{x}[P_2]=r_\beta \mathbf{x}[P_2],
\end{equation}
while the rate of the transition $\beta^{P_2\rightarrow P_3}$ is
\begin{equation}\label{eq:Ch3-f(x,beta)-2}
f(\mathbf{x},\beta^{P_2\rightarrow
P_3})=\frac{r_\beta'}{r_\beta+r_\beta'}
(r_\beta+r_\beta')\mathbf{x}[P_2]=r_\beta' \mathbf{x}[P_2].
\end{equation}

\par In Model~\ref{model:Ch3-P3Q2}, $\alpha$ is a shared activity with
three local rates: $r_\alpha, r_\alpha'$ and $r_\alpha''$. The
apparent rate of $\alpha$ in $P_1$ is
$(r_\alpha'+r_\alpha'')\mathbf{x}[P_1]$, while in $Q_1$ it is
$r_\alpha \mathbf{x}[Q_1]$. According to the PEPA semantics, the
apparent rate of a synchronised activity is the minimum of the
apparent rates of the cooperating components. So the apparent rate
of $\alpha$ as a synchronisation activity is
$\min\{(r_\alpha'+r_\alpha'')\mathbf{x}[P_1], r_\alpha
\mathbf{x}[Q_1]\}$.  After firing $\alpha$, $P_1$ becomes  either
$P_2$ or $P_3$, with the probabilities
$\frac{r_\alpha'}{r_\alpha'+r_\alpha''}$ and
$\frac{r_\alpha''}{r_\alpha'+r_\alpha''}$ respectively.
Simultaneously,  $Q_1$ becomes $Q_2$ with the probability $1$. So
the rate function of transition $(P_1\rightarrow P_2, Q_1\rightarrow
Q_2)$, represented by $f(\mathbf{x},\alpha^{(P_1\rightarrow
P_2,Q_1\rightarrow Q_2)})$, is
\begin{equation}\label{eq:Ch3-f(x,alpha)-1}
\begin{split}
&f(\mathbf{x},\alpha^{(P_1\rightarrow P_2,Q_1\rightarrow
Q_2)})\\=&\frac{r_\alpha'}{r_\alpha'+r_\alpha''}
\min\{(r_\alpha'+r_\alpha'')\mathbf{x}[P_1], r_\alpha
\mathbf{x}[Q_1] \}.
\end{split}
\end{equation}
Similarly,
\begin{equation}\label{eq:Ch3-f(x,alpha)-2}
\begin{split}
&f(\mathbf{x},\alpha^{(P_1\rightarrow P_3,Q_1\rightarrow Q_2)})\\
=&\frac{r_\alpha''}{r_\alpha'+r_\alpha''}
\min\{(r_\alpha'+r_\alpha'')\mathbf{x}[P_1], r_\alpha
\mathbf{x}[Q_1] \}.
\end{split}
\end{equation}

\par The above discussion about the simple example should help the reader
to understand the definition of transition rate function for general
PEPA models, which is presented in the next subsection.

\subsection{Definitions of transition rate function}

\par In a PEPA model, as we have mentioned, we may rewrite any
$U\stackrel{(l,r)}{\longrightarrow}V$ as
$U\stackrel{(l,r_l^{U\rightarrow V})}{\longrightarrow}V$, where $r$
is denoted by $r_l^{U\rightarrow V}$.
 The transition rate functions of general PEPA models are defined below.
We first give the definition of the apparent rate of an activity in
a local derivative.

\begin{definition}(\textbf{{Apparent Rate of $l$ in $U$}})\label{def:DingApparentRate}
 Suppose $l$ is an activity of a PEPA model and $U$ is a local derivative
 enabling $l$ (i.e. $U\in \mathrm{pre}(l)$). Let $\mathrm{post}(U,l)$ be the set of all the local
derivatives derived from $U$ by firing $l$, i.e. $
      \mathrm{post}(U,l)=\{V\mid U\stackrel{(l,r_l^{U\rightarrow V})}{\longrightarrow}V\}.
$ Let
\begin{equation}\label{eq:r_l^U}
       r_l(U)=\sum_{V\in \mathrm{post}(U,l)}r_l^{U\rightarrow V}.
\end{equation}
The \emph{apparent rate} of $l$ in $U$ in  state $\mathbf{x}$,
denoted by $r_l(\mathbf{x},U)$, is defined as
\begin{equation}\label{eq:r_l(U)}
         r_l(\mathbf{x},U)=\mathbf{x}[U]r_l(U).%=\sum_{V\in post(U,l)}r_l^{U\rightarrow V},
\end{equation}
\end{definition}
The above definition is used to define the following transition rate
function.
\begin{definition}(\textbf{{Transition Rate Function}})
\label{def:TransitionRateFunction} Suppose $l$ is an activity of a
PEPA model and
 $\mathbf{x}$ denotes a state vector.

\begin{enumerate}
  \item  If $l$ is individual, then for each
  $U\stackrel{(l,r^{U\rightarrow V})}{\longrightarrow}V$,
the transition rate function of labelled activity $l^{U\rightarrow
V}$ in state $\mathbf{x}$ is defined as
\begin{equation}\label{eq:Ch3-RateFunction1}
     f(\mathbf{x}, l^{U\rightarrow V})=\mathbf{x}[U]r_l^{U\rightarrow V}.
\end{equation}

\item If $l$ is synchronised, with
$\mathrm{pre}(l)=\{U_1,U_2,\cdots,U_k\}$, then for each
$(V_1,V_2,\cdots,V_k)$ in $\mathrm{post}(U_1,l)\times
\mathrm{post}(U_2,l)\times\cdots\times \mathrm{post}(U_k,l),$
 let $w=(U_1\rightarrow V_1,U_2\rightarrow
V_2,\cdots,U_k\rightarrow V_k)$. Then the transition rate function
of labelled activity $l^{w}$ in state $\mathbf{x}$ is defined as
\begin{equation*}
f(\mathbf{x}, l^w)=\left(\prod_{i=1}^k\frac{r_l^{U_i\rightarrow
V_i}}{r_l(U_i)}\right)\min_{i\in\{1,\cdots,k\}}\{r_l(\mathbf{x},U_i)\},
\end{equation*}
where $r_l(\mathbf{x},U_i)=\mathbf{x}[U_i]r_l(U_i)$ is the apparent
rate of $l$ in $U_i$ in state $\mathbf{x}$. So
\begin{equation}\label{eq:Ch3-RateFunction20}
f(\mathbf{x}, l^w)=\left(\prod_{i=1}^k\frac{r_l^{U_i\rightarrow
V_i}}{r_l(U_i)}\right)\min_{i\in\{1,\cdots,k\}}\{\mathbf{x}[U_i]r_l(U_i)\}.
\end{equation}
\end{enumerate}
\end{definition}

\begin{remark}\label{remark:Ch3-Top*0=0}
Definition~\ref{def:TransitionRateFunction} accommodates the passive
or unspecified rate $\infty$. If some $r^{U\rightarrow V}_l$ are
$\infty$, then the relevant calculation in the rate functions
(\ref{eq:Ch3-RateFunction1}) and (\ref{eq:Ch3-RateFunction20}) can
be made  according to the inequalities and equations that define the
comparison and manipulation of unspecified activity rates (see
\cite{Jane1}). Moreover, we assume that $0\cdot\infty=0$. So the
terms such as ``$\min\{A\infty, rB\}$'' are interpreted
as~\cite{WormAttacks}:
\begin{equation*}
    \min\{A\infty, rB\}=\left\{\begin{array}{cc}
                                 rB, & A>0, \\
                                 0, & A=0.
                               \end{array}
    \right.
\end{equation*}
\end{remark}

\par The definition of the transition rate function is consistent with
the semantics of PEPA:
\begin{proposition}\label{proposition:Ch3-RateFunctionConsistency}
The transition rate function in
Definition~\ref{def:TransitionRateFunction} is consistent with the
operational semantics of PEPA.
\end{proposition}
The proof is easy and omitted here. Since both the structural and
timing information has been captured in the defined numerical
representation schema, PEPA models can be therefore recovered from
its representation schema. In addition, it is also easy to find that
the transition rate function has the following  homogenous property.
\begin{proposition}\label{proposition:Ch3-HomogeousAndLipschitz}
The transition rate function $f(\mathbf{x},l)$ satisfies that for
any $H>0$, $Hf(\mathbf{x}/H,l)=f(\mathbf{x},l)$.
\end{proposition}
This property will identify the CTMCs underlying a PEPA model to be
density dependent (see Theorem~\ref{thm:DensityDependentMCfromPEPA}
in the next section).

\subsection{Algorithm for deriving activity matrix and transition rate functions}
This subsection presents an algorithm for automatically deriving the
activity matrix and transition rate functions from any PEPA model,
see Algorithm~\ref{alg:Ch3-DeriveAlgorithm}. The lines 3-12 of
Algorithm~\ref{alg:Ch3-DeriveAlgorithm} deal with individual
activities while lines $13-32$ deal with shared activities. The
calculation methods in this algorithm are essentially the embodiment
of the definitions of labelled activity and apparent rate as well as
transition rate function. So we do not give more introduction to
this algorithm.

\begin{algorithm*}[htbp]
\caption{Derive activity matrix and transition rate functions from a
general PEPA model } \label{alg:Ch3-DeriveAlgorithm}
\begin{algorithmic}[1]

\STATE $\mathcal{A}_{\mathrm{label}}=\emptyset$; $\mathcal{D}$ is
the set of all local derivatives

\FORALL{activity $l\in \mathcal{A}$}
     \IF{$l$ is an independent activity}
          \FORALL{local derivatives $U,V\in \mathcal{D}$}
                \IF{$U\stackrel{(l,r)}{\longrightarrow}V$}
                         \STATE $\mathcal{A}_{\mathrm{label}}=\mathcal{A}_{\mathrm{label}}
                            \cup\{l^{U\rightarrow V}\}$
                            \quad\quad // Label $l$ as $l^{U\rightarrow V}$
                         \STATE // Form a corresponding column of the activity
                         matrix and the rate function
                        \STATE  $
                                M_a(d, l^{U\rightarrow V})=\left\{
                                               \begin{array}{ll}
                                                 -1, & d=U \\
                                                 1, & d=V \\
                                                 0, & otherwise
                                               \end{array}
                              \right.
                              $
                         %\STATE // Form the rate function of $l^{U\rightarrow V}$
                         \STATE $
                                 f(\mathbf{x}, l^{U\rightarrow V})=r\mathbf{x}[U]
                         $

               \ENDIF
          \ENDFOR
    \ENDIF

\IF{$l$ is a synchronised activity}

          \STATE $\mathrm{pre}(l)=\emptyset, \mathrm{post}(U,l)=\emptyset,\; \forall U\in \mathcal{D}$

          \FORALL{local derivatives $U, V\in\mathcal{D}$}
                \IF{$U\stackrel{(l,r)}{\longrightarrow}V$}
                         \STATE $\mathrm{pre}(l)=\mathrm{pre}(l)\cup\{U\}$
                         \STATE $\mathrm{post}(U,l)=\mathrm{post}(U,l)\cup\{V\}$
                         \STATE $r_l^{U\rightarrow V}=r$
               \ENDIF
          \ENDFOR

\STATE Denote
$\mathrm{pre}(l)=\{\mathrm{pre}(l)[1],\mathrm{pre}(l)[2],\cdots,\mathrm{pre}(l)[k]\}$,
where $k=\#\mathrm{pre}(l)$

\FOR{$i=1\dots k$}
   \STATE $r_l(\mathrm{pre}(l)[i])=\sum\limits_{V\in
\mathrm{post}({\mathrm{pre}(l)[i]},l)}
r_l^{\mathrm{pre}(l)[i]\rightarrow V}$ \ENDFOR

 \STATE $K(l)=\mathrm{post}({\mathrm{pre}(l)[1],l})\times
\mathrm{post}({\mathrm{pre}(l)[2]},l)\times\cdots\times
\mathrm{post}({\mathrm{pre}(l)[k]},l)$

\FORALL{$(V_1,V_2,\cdots,V_k)\in K(l)$}

   \STATE $w=(\mathrm{pre}(l)[1]\rightarrow V_1,\mathrm{pre}(l)[2]\rightarrow V_2,\cdots,\mathrm{pre}(l)[k]\rightarrow V_k)$
   \STATE  $\mathcal{A}_{\mathrm{label}}=\mathcal{A}_{\mathrm{label}}
                            \cup\{l^{w}\}$
                            \quad\quad // Label $l$ as $l^{w}$
   \STATE // Form a column of $M_a$ and the rate function
   corresponding to $l^{w}$
   %\STATE
    $$
                              M_a(d, l^w)=\left\{
                                               \begin{array}{ll}
                                                 -1, & d\in \mathrm{pre}(l) \\
                                                 1, & d\in\{V_1,V_2,\cdots,V_k\} \\
                                                 0, & otherwise
                                               \end{array}
                              \right.
                              $$
          \begin{equation*}
          \begin{split}
              &f(\mathbf{x},l^w)=\left(\prod_{i=1}^k
                        \frac{r_l^{\mathrm{pre}(l)[i]\rightarrow V_i}}{r_l(\mathrm{pre}(l)[i])}\right)
                                \min_{i\in\{1,\cdots,k\}}\{r_l(\mathrm{pre}(l)[i])\mathbf{x}[\mathrm{pre}(l)[i]]\}
         \end{split}
         \end{equation*}

 \ENDFOR
\ENDIF

\newpage

\ENDFOR

\STATE Output $\mathcal{A}_{\mathrm{label}}$; $M_a$;
$f(\mathbf{x},l)\; (\forall l\in \mathcal{A}_{\mathrm{label}} )$.
\end{algorithmic}
\end{algorithm*}

\section{Computational approaches for PEPA}
\label{section:Ch3-AssociatedMethods}

As a model being represented numerically, efficient techniques such
as stochastic simulation and fluid approximation can be  can be
directly utilised to analyse the model. This section briefly
introduces these approaches as well as technical foundations for
employing them in the context of PEPA.

\subsection{Place/Transition structure in PEPA models}

Whilst the focus of stochastic process algebras has understandably
been primarily quantitative analysis, qualitative analysis can also
provide valuable insight into the behaviour of a system. In
contrast, in Petri net modelling there are well-established
techniques of structural analysis
\cite{pt-Linear-algebraic-techniques,pt_LinearAlgebricProgramming,pt_LogicalPropertiesPT}.
This subsection shows how the new representation schema helps to
manifest the P/T structure underlying PEPA models, and makes it
possible to readily adapt structural analysis techniques for Petri
nets to PEPA\@. First, the relevant definitions are given below.

\begin{definition}(\textbf{P/T net, Marking, P/T system}, \cite{pt_LogicalPropertiesPT})
\label{def:ChStru-P/Tstructure}
\begin{enumerate}
  \item A \emph{Place/Transition net} (P/T net) is a structure
$\mathcal{N}=(P,T,\mathbf{Pre, Post})$ where: $P$ and $T$ are the
sets of \emph{places} and \emph{transitions} respectively;
$\mathbf{Pre}$ and $\mathbf{Post}$ are the $|P|\times |T|$ sized,
natural valued, \emph{incidence matrices}.

  \item A \emph{marking} is a vector $\mathbf{m}:
P\rightarrow \mathbb{N}$ that assigns to each place of a P/T net a
nonnegative integer (number of tokens).

  \item A \emph{P/T system} is a pair
$\mathcal{S}=\langle \mathcal{N},\mathbf{m_0} \rangle$: a net
$\mathcal{N}$ with an initial marking $\mathbf{m_0}$.
\end{enumerate}
\par
\end{definition}

By Definition~\ref{def:ChStru-P/Tstructure}, it is easy and direct
to verify
\begin{theorem}\label{theorem:ChStru-P/TinPEPA}
There is a P/T system underlying any PEPA model, that is
$\langle\mathcal{N},\mathbf{m_0}\rangle$, where $\mathbf{m_0}$ is
the starting state;
$\mathcal{N}=\left(\mathcal{D},\mathcal{A}_{\small
\mbox{label}},\mathbf{C^{\mbox{Pre}}, C^{\mbox{Post}}}\right)$ is
P/T net: where $\mathcal{D}$ is the set of local derivatives,
$\mathcal{A}_{\small \mbox{label}}$ is the labelled activity set;
$\mathbf{C^{\mbox{Pre}}}$ and  $\mathbf{C^{\mbox{Post}}}$ are the
pre and post activity matrices respectively. Moreover, each state
$\mathbf{m}$ of  the  PEPA model is a marking.
\end{theorem}

Based on the P/T structure underlying PEPA models  and the theories
developed for P/T nets, several powerful techniques and approaches
for structural analysis of PEPA were established
in~\cite{Structural-Analysis-PEPA}. For instance, the authors gave a
method of deriving and storing the state space which avoids the
problems associated with populations of components, and an approach
to find invariants which can be used to qualitatively reason about
systems. Moreover, a structure-based deadlock-checking algorithm was
proposed, which can avoid the state space explosion problem.

\subsection{Stochastic simulation of PEPA models}

By solving the global balance equations associated with the
infinitesimal generator of the CTMC underlying a PEPA model, the
steady-state probability distribution can be obtained, from which
performance measures can be derived. According to the original
definition of the PEPA language in which each instance of the same
component type is considered distinctly, the size of the state space
of this original CTMC may increase exponentially with the number of
components. By adopting the numerical vector form to represent the
system state which results in the aggregated CTMC\@, the size of the
state space can thus be significantly reduced, as
Proposition~\ref{proposition:Ch3-NumericalVector} shows, together
with the computational complexity of deriving the performance by
solving the corresponding global balance equations since, the
dimension of the infinitesimal generator matrix is the square of the
size of the state space.

\par  Unless otherwise stated, hereafter the CTMC underlying a PEPA model
refers to the aggregated CTMC, and the state of a model or a system
is considered in the sense of aggregation. If the size of the state
space is too large, it is not feasible to calculate the steady-state
distribution and thus to get a performance measure $R$, which is
usually expressed as $R=\sum_{\mathbf{s}\in S}\rho
(\mathbf{s})\mathbf{\pi} (\mathbf{s})$ where $\rho$ and
$\mathbf{\pi}$ defined on the state space $S$ are the reward
function and the steady-state probability distribution respectively.
An alternative widely-used way to obtained performance is stochastic
simulation.
% In fact, Gillespie's
%stochastic simulation algorithm (SSA)~\cite{Gillespie1976} has
%already been implemented in the PEPA Eclipse
%plug-in~\cite{PEPA-Plug-in}, a tool supporting PEPA\@. The Gillespie
%algorithm has been widely applied to model and simulate biochemical
%reactions~\cite{STOCKS-Gillespie-Algorithm}. This method exploits
%the fact that the duration from one transition (or reaction) to the
%next satisfies an exponential distribution with the reciprocal of
%total transition rate as the mean, and assumes that the transition
%rate is dependent on the state~\cite{SSA-Matlab}. In the context of
%PEPA, this algorithm can be directly constructed based on our
%numerical representation schema.

\par As discussed
previously, a transition between states, namely from $\mathbf{x}$ to
$\mathbf{x}+l$, is represented by a transition vector $l$
corresponding to the labelled activity $l$ (for convenience,
hereafter each pair of transition vectors and corresponding labelled
activities shares the same notation). The rate of the transition $l$
in state $\mathbf{x}$ is specified by the transition rate function
$f(\mathbf{x},l)$. That is, $
   \mathbf{x}\mathop{\longrightarrow}\limits^{(l,f(\mathbf{x},l))}\mathbf{x}+l.
$ Given a starting state $\mathbf{x}_0$, the transition chain
corresponding to a firing sequence $l_0,l_1,\cdots,l,\cdots$ is
\begin{equation*}
\begin{split}
   &\mathbf{x}_0\mathop{\longrightarrow}\limits^{(l_0,f(\mathbf{x}_0,l_0))}
   \mathbf{x}_0+l_0
   \mathop{\longrightarrow}\limits^{(l_1,f(\mathbf{x}_0+l_0,l_1))}
   (\mathbf{x}_0+l_0)+l_1\mathop{\longrightarrow}\limits^{\cdots}\\&\cdots
   \mathop{\longrightarrow}
   \limits^{\cdots}\mathbf{x}\mathop{\longrightarrow}\limits^{(l,f(\mathbf{x},l))}
   \mathbf{x}+l\mathop{\longrightarrow}\limits^{\cdots}\cdots.
\end{split}
\end{equation*}
The above sequence can be considered as one path or realisation of a
simulation of the aggregated CTMC, if the enabled activity at each
state is chosen stochastically, i.e.\ is chosen through the approach
of sampling. After a long time, the steady-state of the system is
assumed to be achieved. Hence the average performance
$R=\sum_{\mathbf{s}\in S}\rho (\mathbf{s})\mathbf{\pi} (\mathbf{s})$
can be calculated.

As one benefit of our numerical representation schema, it provides a
good platform for directly and conveniently simulating the CTMC for
PEPA, see Algorithm~\ref{alg:StochasticSimulation}. In
Algorithm~\ref{alg:StochasticSimulation},
 the states of a PEPA model are represented as
numerical vector forms, and the  rates between those states are
specified by the transition rate functions which only depend on the
transition type (i.e.\ labelled activity) and the current state. In
this algorithm, the generated time $\tau$ in each iteration can be
regarded as having been drawn from an exponential distribution with
the mean $\displaystyle\frac{1}{f(\mathbf{x})}$ (see Example 2.3
in~\cite{StochasticSimulation-book}, page 38). That is, Line $9$ in
Algorithm~\ref{alg:StochasticSimulation} is in fact expressing:
{``generate $\tau$ from an exponential distribution with the mean
$\displaystyle\frac{1}{f(\mathbf{x})}$''.} Line $10$ determines
which transition will be chosen, and consequently determines the
next state that the system will transition into. Therefore, this
algorithm is essentially to simulate the CTMC underlying a PEPA
model.

% we
%emphasise that any general performance measure can be derived from
%PEPA models through this kind of simulation, particularly those that
%cannot be obtained by other approach (e.g.\ the fluid approximation
%approach).

% which has already been pointed out in~\cite{Gillespie1976}.

\begin{algorithm*}[t]
\caption{Simulation algorithm for deriving general performance
measures from PEPA model} \label{alg:StochasticSimulation}

\begin{algorithmic}[1]

\STATE //Initialisation

\STATE starting state $\mathbf{x}$; labelled activity set
$\mathcal{A}_{\mathrm{label}}=\{l_1,l_2,\cdots,l_m\}$; activity
matrix; transition rate function $f$\\%[4pt]

\STATE  reward
function $\rho$; $\mathrm{PerMeasure}=0$\\%[4pt]

\WHILE{stop condition not satisfied}

\STATE //Sampling\\%[4pt]

\STATE compute the transition rate function
$\displaystyle{f(\mathbf{x},l_j)}$, $j=1,2,\cdots,m$\\%[4pt]

\STATE $f(\mathbf{x})=\sum_{j=1}^m f(\mathbf{x},l_j)$\\%[4pt]

\STATE generate uniform random numbers $r_1,r_2$ on $[0,1]$\\%[4pt]

\STATE compute $\tau=-(1/f(\mathbf{x}))\ln r_1$\\%[4pt]

\STATE find $\mu$ such that $\sum_{j=1}^{\mu-1}f(\mathbf{x},l_j)\leq
r_2f(\mathbf{x}) <\sum_{j=1}^{\mu}f(\mathbf{x},l_j)$\\%[4pt]

\STATE //Updating\\%[4pt]

\STATE
$\mathrm{PerMeasure}=\mathrm{PerMeasure}+\rho(\mathbf{x})\times
\tau$ \quad // Accumulate performance measure\footnotemark\\%[4pt]

\STATE $t=t+\tau$ \quad //Accumulate time\\%[4pt]

\STATE $\mathbf{x}=\mathbf{x}+l_{\mu}$ \quad // Update state vector
of system\\%[4pt]

 \ENDWHILE\\%[4pt]

\STATE Output performance:
$\frac{\mathrm{PerMeasure}}{t}$\\%[4pt]
\end{algorithmic}
\end{algorithm*}
\footnotetext{In practise, in order to decrease the computational
cost, we should not calculate the performance until after a warm up
period so that the effects of the initial state bias can be
considered to be negligible.}

The choices for stopping the algorithm include a given large time,
or the absolute or relative error of two continued iterations being
small enough (since the output performance converges as time goes to
infinity as the following
Theorem~\ref{thm:Ch7-Performance-Convergence} states). Now we prove
the convergence of the performance calculated using the algorithm.
We need the following theorem.
\begin{theorem}(Theorem~3.8.1,
\cite{Norris-MarkovChains})\label{thm:Ergodic-CTMC-Convergence} If
$X(t)$ is an irreducible and positive recurrent CTMC with the state
space $S$ and the unique invariant distribution $\mathbf{\pi}$, then
\begin{equation}
  \mathrm{Pr}\left(
\frac{1}{t}\int_{0}^t1_{\{X_{z}=\mathbf{s}\}}\mathrm{d}z\rightarrow
\mathbf{\pi}(\mathbf{s})\mbox{ as }t\rightarrow \infinity \right)=1.
\end{equation}
Moreover, for any bounded function $\rho: S\rightarrow \mathbb{R}$,
we have
\begin{equation}
  \mathrm{Pr}\left(
\frac{1}{t}\int_{0}^t\rho(X_{z})\mathrm{d}z\rightarrow
E[\rho(X)]\mbox{ as }t\rightarrow \infinity \right)=1.
\end{equation}
where $E[\rho(X)]=\sum_{\mathbf{s}\in
S}\rho(\mathbf{s})\mathbf{\pi}(\mathbf{s})$.
\end{theorem}
Here is our conclusion (we assume the CTMCs underlying PEPA models
to be irreducible and positive recurrent):
\begin{theorem}\label{thm:Ch7-Performance-Convergence}
The performance measure calculated according to
Algorithm~\ref{alg:StochasticSimulation} converges as time goes to
infinity, that is, $\lim_{t\rightarrow
\infinity}\frac{\mathrm{PerMeasure}}{t}=E[\rho(X)]$.
\end{theorem}

\begin{proof}
Assume that $n-1$ iterations have been finished and the time has
accumulated to $t_{n-1}$. Suppose the current one is the $n$-th
iteration and  $\tau$ is the generated time in this iteration. After
the $n$-th iteration is finished, the accumulated time will be
updated to $t_n=t_{n-1}+\tau$. During the $\tau$ time interval, the
simulated CTMC stays in the state $\mathbf{x}$, that is,
$X_z=\mathbf{x}, z\in[t_{n-1},t_n)$. So,
\begin{equation*}
\begin{split}
     &\rho(\mathbf{x})\times \tau\\
     =&\rho(\mathbf{x})\int_{t_{n-1}}^{t_{n}}
\mathrm{d}z=\int_{t_{n-1}}^{t_{n}} \rho(\mathbf{x})\mathrm{d}z
   =\int_{t_{n-1}}^{t_{n}}\rho(X_{z})\mathrm{d}z.
\end{split}
\end{equation*}
Therefore, after this $n$-th iteration, $\mathrm{PerMeasure}$ will
be accumulated to $=\int_{0}^{t_{n}}\rho(X_{z})\mathrm{d}z$ and
$$\frac{\mathrm{PerMeasure}}{t_n}
=\frac{1}{t_n}\int_{0}^{t_{n}}\rho(X_{z})\mathrm{d}z.
$$
According to Theorem~\ref{thm:Ergodic-CTMC-Convergence},
$\displaystyle\frac{1}{t_n}\int_{0}^{t_{n}}\rho(X_{z})\mathrm{d}z$
tends to $ E[\rho(X)]$ as $t_n$ tends to infinity. So the
performance obtained through
Algorithm~\ref{alg:StochasticSimulation} converges to $E[\rho(X)]$
as the simulation time goes to infinity.
\end{proof}

\par Performance metrics,  such as activity throughput of an
activity and capacity utilisation of a local derivative that are
discussed in~\cite{MircoJie-Fluid-Reward}, can be derived through
this algorithm by choosing appropriate reward functions.

\subsection{Fluid approximation of PEPA models}
The weakness of the simulation method is its high computational
cost, which makes it not suitable for real-time performance
monitoring or prediction. Recently, a novel approach to get
performance measures from PEPA models has been proposed
in~\cite{Jane2} and subsequently expanded
in~\cite{Mirco-Fluid-Semantics} and~\cite{JieThesis}, making a
continuous state space approximation as a set of ordinary
differential equations (ODEs). In this section, we present a mapping
semantics for this approaches, which is based on the numerical
representation schema. In addition, a theoretical justification of
this approach, mainly in terms of its consistency with the CTMCs,
will be discussed.

%\subsubsection{Mapping semantics}
In our representation schema, the transition rate $f(\mathbf{x},l)$
reflects the intensity of the transition from state $\mathbf{x}$ to
state $\mathbf{x}+l$. The state space is inherently discrete with
the entries within the numerical vector form always being
non-negative integers and always being incremented or decremented in
steps of one. As pointed out in~\cite{Jane2}, when the numbers of
components are large these steps are relatively small and we can
approximate the behaviour by considering the movement between states
to be continuous, rather than occurring in discontinuous jumps. In
fact, let us consider the evolution of the numerical state vector.
Denote the state at time $t$ by $\mathbf{x}(t)$. In a short time
$\Delta t$, the change to the vector $\mathbf{x}(t)$ will be
\begin{equation*}
\begin{split}
       \mathbf{x}(\cdot,t+\Delta t)-\mathbf{x}(\cdot,t)
       =\Delta t\sum_{l\in\mathcal{A}_{\mathrm{label}}}lf(\mathbf{x}(\cdot,t),l).
\end{split}
\end{equation*}
Dividing by $\Delta t$ and taking the limit, $\Delta t\rightarrow
0$, we obtain  a set of ODEs:
\begin{equation}\label{eq:ChFP-DingDerivedODEs}
   \frac{\mathrm{d}\mathbf{x}}{\mathrm{d}t}=\sum_{l\in\mathcal{A}_{\mathrm{label}}}lf(\mathbf{x},l).
\end{equation}

\par Once the  activity matrix and the transition rate functions are
generated, the ODEs are immediately available. All of them can be
obtained automatically by Algorithm~\ref{alg:Ch3-DeriveAlgorithm}.

For an arbitrary CTMC, the evolution of probabilities distributed on
each state can be described using linear ODEs
(see~\cite{QueueingNetworks}, page 52). For example, for the
aggregated CTMC underlying a PEPA model, the corresponding
differential equations are
\begin{equation}\label{eq:ChFP-KolmogorovProbDistr}
    \frac{\mathrm{d}\mathbf{\pi}}{\mathrm{d}t}=Q^T\mathbf{\pi},
\end{equation}
where each entry of $\mathbf{\pi}(t)$ represents the probability of
the system being in  each state at time $t$, and $Q$ is an
infinitesimal generator matrix corresponding to the CTMC. Clearly,
the dimension of the coefficient matrix $Q$ is the square of the
size of the state space, which increases with the number of
components.

\par The scale of (\ref{eq:ChFP-KolmogorovProbDistr}), i.e.\ the
number of the ODEs, depends on the size of the state space, so it
suffers from the state sapce explosion problem. In contrast, the
ODEs (\ref{eq:ChFP-DingDerivedODEs}) reflect the evolution of the
population of the components in \emph{each local derivative}, so the
scale is only determined by the number of local derivatives and is
unaffected by the size of the state space. Therefore, it avoids the
explosion problem. But the price paid is that the ODEs
(\ref{eq:ChFP-DingDerivedODEs}) are generally nonlinear due to
synchronisations, whereas (\ref{eq:ChFP-KolmogorovProbDistr}) is
linear.

%\subsubsection{Consistency for ODEs}
\par This paper emphasises the consistency between the fluid
approximation and the aggregated CTMC\@. Obviously, the CTMC depends
on the starting state of the given PEPA model. By altering the
population of components presented in the model, which can be done
by varying the initial states, we may get a sequence of aggregated
CTMCs. Moreover, the homogenous property that the transition rate
function satisfies, indicated in
Proposition~\ref{proposition:Ch3-HomogeousAndLipschitz}, identifies
the aggregated CTMC to be \emph{density dependent}.
\begin{definition}(\cite{Kurtz}).
A family of CTMCs $\{X_n\}_n$ is called \emph{density dependent} if
and only if there exists a continuous function
$f(\mathbf{x},l),\;\mathbf{x}\in \mathbb{R}^d,\;l\in \mathbb{Z}^d$,
such that the infinitesimal generators of $X_n$ are given by:
$$
           q^{(n)}_{\mathbf{x},\mathbf{x}+l}=nf(\mathbf{x}/n,l),\quad l\neq 0,
$$
where $q^{(n)}_{\mathbf{x},\mathbf{x}+l}$ denotes an entry of the
infinitesimal generator of $X_n$, $\mathbf{x}$ a numerical state
vector and $l$ a transition vector.
\end{definition}

This allows us to immediately conclude the following conclusion:
\begin{theorem}\label{thm:DensityDependentMCfromPEPA}(\cite{JieThesis,Mirco-Fluid-Semantics})
 Let $\{X_n\}$ be a sequence of
aggregated CTMCs generated from a given PEPA model (by scaling the
initial state), then $\{X_n\}$ is density dependent.
\end{theorem}

\par Since both ODEs and density dependent CTMCs
can be derived from the same PEPA model through the same activity
matrix and transition rate functions, it is natural to believe some
kind of consistency between them. In fact, according to Kurtz's
theorem~\cite{KurtzBook}, the complete solution of some ODEs can be
the limit of a sequence of Markov chains. Such consistency in the
context of PEPA has been previously illustrated for a particular
PEPA model in~\cite{Jane4}, and subsequently generalised to general
models in~\cite{Mirco-Fluid-Semantics} and~\cite{JieThesis}. The
result presented below is extracted from~\cite{JieThesis}, in which
the convergence is in the sense of almost surely rather than
probabilistically as in~\cite{Jane4}
and~\cite{Mirco-Fluid-Semantics}.

\begin{theorem}\label{thm:ChFP-ODEmeetsAggCTMC}(\cite{JieThesis})
Let $\mathbf{x}(t)$ be the solution of the ODEs
(\ref{eq:ChFP-DingDerivedODEs}) derived from a given PEPA model with
initial condition $\mathbf{x}_0$, and let $\{X_n(t)\}$ be the
density dependent CTMCs with $X_n(0)=n\mathbf{x}_0$.
 Then for any $t>0$,
\begin{equation}
        \lim_{n\rightarrow\infinity}\sup_{u\leq t}\left\|X_n(u)/n-\mathbf{x}(u)\right\|=0\quad \quad a.s.
\end{equation}
\end{theorem}
This theorem justifies the fluid approximation by manifesting the
consistency between this approach and the corresponding CTMCs for a
general PEPA model. Furthermore, if there is no synchronisation
contained in the model then the derived ODEs
(\ref{eq:ChFP-DingDerivedODEs}) becomes linear, and
(\ref{eq:ChFP-KolmogorovProbDistr}) and
(\ref{eq:ChFP-DingDerivedODEs}) coincide except for a constant
factor. Moreover, the fundamental results on the fluid approximation
of PEPA models such as the existence, uniqueness, boundedness and
nonnegativeness of the ODEs' solution, as well as the solution's
asymptotic behaviour, have been obtained. In particular, the
convergence of the ODEs' solution  as time tends to infinity, has
been proved under a condition, which is revealed to relate to some
famous constants of Markov chains such as the spectral gap and the
Log-Sobolev constant. For more details about these stories, please
refer to~\cite{JieDing_ODEsFundamental}. As for performance
derivation via this approach, please
see~\cite{MircoJie-Fluid-Reward}.

\section{Conclusions}

\par In this paper we have demonstrated a schema, which bridges the syntactic and numerical
representation, as well as the local definition and global analysis
for a PEPA model. Computational approaches and associated algorithms
developed based on the schema have been presented, which can help to
relieve the state space explosion problem for large scale models.
For other stochastic process algebras, similar numerical
representation schema can be established and expected to benefit
relevant performance modelling.

\section*{Acknowledgment}
 Part of this research has been carried out while Jie Ding was at
the University of Edinburgh as a PhD student funded by the Mobile
VCE (www.mobilevce.com), with the School of Informatics and the
School of Engineering.

{
\bibliographystyle{IEEEtr}
\bibliography{JieDing_September2010-2}
}

\end{document}